\documentclass[prd,twocolumn,
showpacs,preprintnumbers,amsmath,amssymb,
a4paper,nofootinbib,longbibliography]{revtex4-2}

\usepackage{setspace}
\usepackage{nicefrac}
\usepackage{esint}

\usepackage{empheq}
\usepackage[most]{tcolorbox}

\usepackage{natbib}
\usepackage{amssymb,amsbsy,amsmath,amsfonts}
\usepackage{graphicx}
\usepackage{epsf,epsfig,float,latexsym,amsthm,fancyhdr,rotating}
\usepackage{graphics,psfrag,tabularx}
\newcolumntype{C}{>{\centering\arraybackslash}X}
\usepackage{slashed}
\usepackage[normalem]{ulem}
\usepackage[colorlinks,citecolor=blue,linktoc=all,linkcolor=cyan,urlcolor=blue]{hyperref}
\usepackage{upgreek}
\usepackage{tikz}
\usepackage{tikz-feynman}
\tikzfeynmanset{compat=1.1.0} 

\tikzfeynmanset{
    doublefermion/.style={
        /tikz/draw=black,
        double distance=1.2pt, 
        line width=0.4pt,
        postaction={draw, line width=0.4pt, white, shorten >=0.4pt, shorten <=0.4pt}
    }
}
 
\usepackage{multirow}

\def\beq{\begin{equation}}
\def\eeq{\end{equation}}
\def\bea{\begin{eqnarray}}
\def\eea{\end{eqnarray}}
\def\beqa{\begin{equation}\begin{array}{l}}
\def\eeqa{\end{array}\end{equation}}
\def\eqlab#1{\label{eq:#1}}

\def\seclab#1{\label{sec:#1}}
\def\eref#1{(\ref{eq:#1})}
\def\Eqref#1{Eq.~(\ref{eq:#1})}

\def\Figref#1{Fig.~\ref{fig:#1}}

\def\secref#1{Sec.~\ref{sec:#1}}

\def\barr{\left(\begin{array}{c}}
\def\earr{\end{array}\right)}
\def\bmat{\left(\begin{array}{cc}}
\def\emat{\end{array}\right)}
\def\al{\alpha}

\def\ga{\gamma}

\def\si{\sigma}

\def\ie{\emph{i.e.}}
\def\eg{\emph{e.g.}}

\def\nn{\nonumber}
\def\dd{\mathrm{d}}

\DeclareMathOperator\arctanh{arctanh}

\DeclareMathOperator\im{Im}

\def\alem{\alpha}

\def\au{a_\mu}
\def\ae{a_e}
\def\aue{a_{\mu-e}}
\def\auhvp{a_\mu^\mathrm{HVP}}
\def\aehvp{a_e^\mathrm{HVP}}
\def\auehvp{a_{\mu-e}^\mathrm{HVP}}

\def\ga{\left(}
\def\dr{\right)}
\usepackage{physics}

\usepackage{xcolor}
\usepackage{bm}
\usepackage{slashed}
\usepackage{graphicx}
\allowdisplaybreaks

\makeatletter
\g@addto@macro\bfseries{\boldmath}
\makeatother

\begin{document}

\author{Siyuan Li}
\author{Vladimir Pascalutsa}
\affiliation{Institut f\"ur Kernphysik,
 Johannes Gutenberg-Universit\"at  Mainz,  D-55128 Mainz, Germany}
 
\author{Maxim Pospelov}
\affiliation{William I.\ Fine Theoretical Physics Institute, University of Minnesota, Minneapolis, MN 55455, United States}
\affiliation{Theoretical Physics Department, CERN, 1 Esplanade des Particules, CH-1211 Geneva 23, Switzerland}

\title{Lepton $g-2$ non-universality of hadronic contributions and a sub-GeV window to New Physics}

\begin{abstract}
We propose the linear combination of the anomalous magnetic moments of the muon and electron, \( a_{\mu-e} \equiv a_\mu - (m_\mu/m_e)^2 a_e \), as a natural low-energy window quantity that may directly probe the current discrepancy between the data-driven and lattice-QCD evaluations of hadronic contributions. 
The rescaling ensures an exact cancellation of the leading short-range ($m_\ell^2$-scaling) effects, thereby improving the UV behavior.
Because entire classes of contributions cancel identically --- not merely in variance --- the combination is predictable in chiral perturbation theory with no free low-energy constant, and it isolates flavor-non-universal New Physics.
The hadronic-vacuum-polarization effect in $a^{\rm HVP}_{\mu-e}$, together with its uncertainty, is reduced as compared to $a^{\rm HVP}_{\mu}$ by $\sim 85\%$. This is promising for tests of New Physics with some degree of flavor non-universality, conditional on significant improvements in the measurements of $a_e$ and $\alpha$.
\end{abstract}

\date{\today}

\maketitle
\begin{small}
\end{small}

\section{Introduction}\label{sec:motiv}

The anomalous magnetic moments of the electron and muon,
$a_e$ and $a_\mu$, are some of the best-measured quantities in physics~\cite{Fan:2022eto,Aguillard:2025fij}.
They also come out as a basic prediction of the Standard Model (SM), thus allowing for its stringent precision test and New-Physics constraints
\cite{Melnikov:2006sr,Jegerlehner:2017gek,Aoyama:2020ynm}. 
The dominant theoretical uncertainty in $a_\mu$ comes from the
leading-order (LO) hadronic vacuum polarization (HVP) contribution, for which lattice-QCD and data-driven determinations exhibit a persistent tension (as per the 2025 White Paper of the `$g-2$ Theory Initiative'~\cite{Aliberti:2025beg}).

Moreover, the datasets used in the data-driven evaluations are not entirely consistent with one another, which has led to the omission of this approach from the current SM value altogether. While lattice-QCD calculations appear to be more robust, further improvements are still needed in the evaluation of isospin-breaking effects and systematic uncertainties, such as discretization errors.

In this paper, we address these issues in a combination of $\au$ and $a_e$ in which many systematic effects cancel out. For an earlier discussion of  using $a_e$ to elucidate the hadronic contributions, see, \eg, Refs.~\cite{Giudice:2012pf,Karshenboim:2021jsc,DiLuzio:2024sps}.

It is clear that the hadronic contributions to $a_e$ and $a_\mu$ are strongly 
correlated, but not exactly on the same footing, since $m_e$ is negligible with respect to hadronic scales, such as  $\Lambda_\mathrm{QCD}$, while $m_\mu$ is not. 
On the scale of the electron Compton wavelength, hadronic effects are therefore short-range: the electron does not resolve them, and $\aehvp$ is given by the slope of the vacuum polarization at zero momentum times $m_e^2$ [see \Eqref{ae_slope} below]. The muon, whose mass is comparable to the pion mass, partially resolves the hadronic structure. This does not mean that different energy scales are weighted differently: the hadronic contributions to both are dominated by the same low-energy (long-distance) region, with very similar relative contributions from the different energy scales~\cite{Keshavarzi:2019abf}. Contributions from scales well above $m_\mu$ are proportional to $m_\ell^2$ for both leptons and therefore cancel in the rescaled difference introduced below, which is thus insensitive to hadronic contributions above about 1~GeV.
If $a_e$ is to be utilized for constraining the
hadronic sector, we are inevitably led to consider the (rescaled) difference:\footnote{Alternatively, one can use:
$$a_{e-\mu}
\equiv
a_e - \frac{m_e^2}{m_\mu^2} a_\mu, $$
which may be advantageous when looking at the entire $a_\ell$, not just the hadronic contributions. Also,  $a_{e-\mu}$ is positive-definite whereas
$\aue = - (m_\mu/m_e)^2 a_{e-\mu}$ is negative, which may be another reason 
for preferring the former. As long as the mass-ratio uncertainty is negligible,
using one or the other is equivalent.}
\begin{equation}\label{eq:aue_og_def}
a_{\mu-e}
\equiv
a_\mu - \frac{m_\mu^2}{m_e^2} a_e .
\end{equation}
The mass-ratio rescaling ensures that the high-energy effects
cancel out.\footnote{The exact cancellation of the high-energy contributions is the main advantage of this quantity over the previously considered combination \cite{Giusti:2020efo},
$$ R_{e/\mu} \equiv \frac{m_\mu^2}{m_e^2} \frac{a_e}{a_\mu}
= 1- \frac{a_{\mu-e}}{a_\mu},$$
for which they appear via the standalone $a_\mu$.} 

An analogous scaling between the muonic and ordinary atoms is exploited to reduce the uncertainties due to short-distance proton structure effects.
For example, the theory prediction of the muonic-hydrogen 1S hyperfine
transition is enormously improved by using the same transition
in ordinary hydrogen, the famous 21-cm line  
(see, \eg, Ref.~\cite[Sec.~4.3]{Antognini:2022xoo} for a review). 

In the future, given the appropriately improved knowledge of $a_e$ and the fine-structure constant $\alem$ (to isolate the QED contribution), 
$\aue$ will provide a clean handle on the remaining, mostly hadronic, contributions. 
It will complement, but not replace, separate analyses of $\au$ and $\ae$, whose different dependence on energy scales remains essential for clarifying discrepancies and testing the SM.
In the meantime, it may play the role of a natural `window quantity',  which isolates the sub-GeV range of hadronic contributions. 

The concept of window quantities, introduced in the context of lattice calculations~\cite{RBC:2018dos}, is useful for a more focused comparison with the data-driven approach, as well as for complementing lattice results with data-driven input~\cite{Boccaletti:2024guq}. Since then, generalized windows and optimized weight functions have been constructed to project out particular regions of energy or Euclidean time and to minimize uncertainties, a task closely tied to the inverse reconstruction of the spectral density~\cite{Hansen:2019idp}. Those are weights applied to $\au$ alone; $\aue$ is a window of a different kind --- a physical, measurable observable in its own right.

It therefore serves a somewhat different purpose, elucidating the existing discrepancy not only between lattice and data-driven evaluations (seen in the first column of Table~\ref{tab:ref_values}), but also among different data-driven evaluations as their uncertainties improve. As evidenced by the data-driven entry in Table~\ref{tab:ref_values}, both the magnitude of the HVP contribution and, importantly, its uncertainty\footnote{The quoted uncertainties are obtained in the strong-correlation limit: $ \si_{\mu-e} = \big| \si_\mu - (m_\mu/m_e)^2 \si_e \big|$.} are significantly ($\sim  85\% $) reduced  for $a_{\mu-e}$ compared to $a_\mu$.

\begin{table}[t]
\renewcommand{\arraystretch}{2}
\begin{ruledtabular}
\begin{tabular}{c|llr}
LO HVP & $\au\ga 10^{-10}\dr$ & $\ae\ga10^{-14}\dr$ & $\aue \ga 10^{-10}\dr$\\
 \hline
lattice QCD       & $713.2(6.1) $~\cite{Aliberti:2025beg}   & $189.3 (8.2)$~\cite{Budapest-Marseille-Wuppertal:2017okr} & $-96.1(29.0)$
\\
data-driven    &  $692.8 (2.4)$~\cite{Keshavarzi:2019abf}  & $186.10(66)$~\cite{Keshavarzi:2019abf}
& $-102.8(4)$\\
\hline 
discrepancy    & $20.4(6.6)$   &$3.2(8.2)$ & $6.7(29.0)$\\
\end{tabular} 
\end{ruledtabular}
\caption{HVP contribution to $a_\mu,\, a_e$, and $\aue$ combination from lattice QCD and data-driven approach. }
\label{tab:ref_values}
\end{table}

The large lattice-QCD error for $a_e^\mathrm{HVP}$, seen in the table,
can, in the near future, be improved by an order of magnitude to approach the data-driven value in precision. Certainly, a direct
calculation of $\aue^\mathrm{HVP}$ could
achieve a much better precision (see \secref{HVP} for further remarks).

One may also envision calculations of $\aue$ within an effective-field-theory (EFT) framework, such as chiral perturbation theory ($\chi$PT). For an individual lepton anomalous magnetic moment $a_\ell$, $\chi$PT by itself lacks predictive power because an unknown low-energy constant enters already at leading order (see, \eg,~\cite{Knecht:2001qg,Ramsey-Musolf:2002gmi,Melnikov:2006sr}). 
The corresponding counterterm, however, is flavor-universal and multiplies $m_\ell^2$. Schematically,
\beq
a_\ell = \frac{m_\ell^2}{\Lambda^2}\,\Big[\,C(\mu) + b\,\ln\frac{\mu^2}{\Lambda^2} + f(m_\ell)\Big],
\eqlab{LECgeneric}
\eeq
where $\Lambda$ is the relevant hadronic (or New-Physics) mass scale and the dimensionless low-energy constant $C(\mu)$ is the coefficient of the flavor-universal dipole counterterm,
\beq
\mathcal{L}_{\rm ct} = C(\mu)\,\frac{e}{4\Lambda^2}\sum_{\ell} m_\ell\,\bar\psi_\ell\,\sigma_{\mu\nu}F^{\mu\nu}\psi_\ell ,
\eqlab{Lct}
\eeq
which absorbs the UV divergence; the coefficient $b$ of the accompanying logarithm is lepton-independent, and the dimensionless function $f(m_\ell)$ is finite and calculable. Both $C(\mu)$ and the logarithm therefore cancel in the rescaled difference, $\aue = (m_\mu^2/\Lambda^2)\,[f(m_\mu)-f(m_e)]$, which allows for a genuinely parameter-free prediction. In \secref{pi0gamma}, we illustrate this mechanism for the $\pi^0\gamma$ contribution.

Similar considerations apply to the axion-like particles (ALPs), see \eg,~\cite{Marciano:2016yhf,Bauer:2019gfk,Bauer:2020jbp,Galda:2023qjx,Pustyntsev:2025nwm} for recent studies. In such cases, the corresponding divergences are likewise absorbed into local counterterms, whereas suitably constructed combinations such as $\aue$ can isolate the genuinely low-energy, model-independent part of the contribution.
For flavor-universal ALP couplings, the relevant counterterm is the lepton-dipole operator, whose coefficient scales with $m_\ell$ (as, \eg, in minimal flavor violation), so that $a_\ell^{\rm ALP}$ takes the form \eref{LECgeneric} and the counterterm drops out of $\aue$; cf.\ Eq.~(47) of Ref.~\cite{Galda:2023qjx}.

Currently, the $a_{\mu-e}$ quantity is mostly of a methodological interest along the lines of efforts to resolve tensions between data-driven and lattice values for the HVP contributions. In the future, however, both $a_e$ and $\alpha$ measurements can be improved to the point of making $a_{\mu-e}$ an important probe of New Physics scenarios. If experimental accuracy is not a limiting factor but the treatment of the strong interactions is, then it is likely that the constraining power of $a_{\mu -e}$ is going to be greater than that of $a_\mu$ and $a_e$ separately, precisely due to strong cancellations of the hadronic effects in $a_{\mu-e}$ and reduced error on calculation of this quantity.  
We show that in this, distant future perspective, certain classes of beyond the SM (BSM) scenarios can be constrained. In particular, we argue that $\aue$
may provide a deeper reach into the parameter space of sub-GeV New Physics than $a_{\mu}$ or $a_e$ considered separately. We illustrate this in Sec.~\ref{sec:newphysics} using the dark-photon contribution. More broadly, any flavor-non-universal BSM scenario that deviates from the $a^{\rm BSM}_\mu/a^{\rm BSM}_e = m_\mu^2/m_e^2$ scaling can be constrained this way. 

\section{HVP contribution to $\aue$}
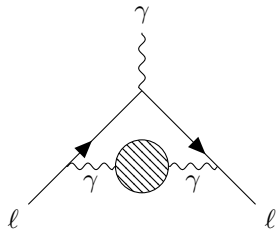
\begin{figure}[htb]
\centering
\begin{tikzpicture}[scale=0.9]
\begin{feynman}

\vertex (a) at (0,1) {\(\gamma\)};
\vertex (b) at (0,0);
\vertex (f1) at (-1,-1);
\vertex (f2) at (1,-1);
\vertex (c) at (-1.7,-1.7){\(\ell\)};
\vertex (d) at (1.7,-1.7){\(\ell\)};
\node[blob, minimum size=7mm]  (f3) at (0,-1);

\diagram* {
(a)-- [boson] (b)-- [plain] (d),
(c)-- [plain] (b),
(f1)-- [boson, edge label'=\(\gamma\)] (f3)-- [boson, edge label'=\(\gamma\)] (f2),
};

\end{feynman}
\end{tikzpicture}
\caption{The leading-order HVP contribution to $(g-2)_\ell$.}
\label{fig:hvp_feynman_diagram}
\end{figure}

\seclab{HVP}
Let us quickly review the properties of $\aue$ for the HVP contribution to lepton $g-2$, seen in Fig.~\ref{fig:hvp_feynman_diagram}. 
Plugging the dispersion relation for vacuum polarization $\Pi(q^2)$ into this diagram, one obtains a familiar expression in terms of its imaginary part~\cite{Bouchiat:1961lbg,PhysRev.168.1620,Lautrup:1968tdb,Gourdin:1969dm}:
\begin{equation}
\label{eq:HVPdisp_og}
a_\ell^{\mathrm{HVP}} = \frac{\alem}{\pi^2}\int\limits_{s_0}^\infty \dd s \, \frac{\im \Pi(s)}{s} 
K_\ell(s) \,,
\end{equation}
with the kernel function given by
\begin{subequations}\label{eq:Ku}
\bea
    K_\ell(s) &=& \frac{m_\ell^2}{s} \int\limits_0^1 \dd x \frac{ x^2 (1-x)}{1-x+x^2( m_\ell^2/s)} \\
    &=& \frac{1}{2} - \frac{s}{m_\ell^2} + \frac{s \ga s- 2 m_\ell^2 \dr}{2m_\ell^4}  \log \frac{s}{m_\ell^2} \nn\\
    && - \ga 2  -\frac{4 s}{m_\ell^2} +\frac{s^2}{m_\ell^4}\dr \nn\\
    && \times \frac{1}{\sqrt{1 - 4 m_\ell^2/s}}\arctanh \sqrt{1 - \frac{4 m_\ell^2}{s} } .
\eea
\end{subequations}

\begin{figure}[t]
    \centering
    \includegraphics[width=0.75\linewidth]{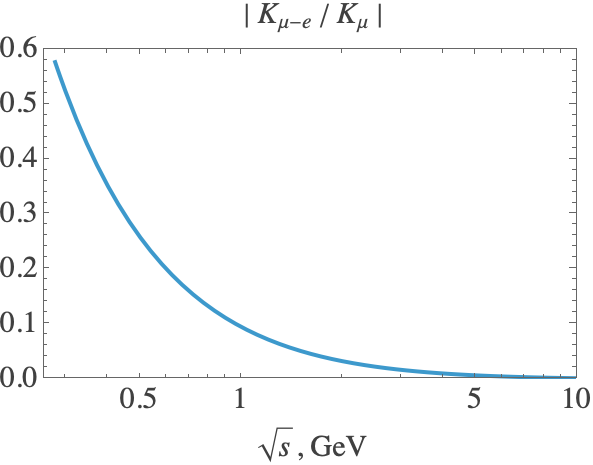}
    \caption{The kernel ratio, $\big| K_{\mu-e}(s)/K_\mu(s)\big|$, as a function of energy. 
    }
    \label{fig:kernel_ratio}
\end{figure}

In the case of HVP, the lowest energy $s_0$ is associated with a pion-production threshold. Hence,
for the electron, we may --- to an excellent approximation\footnote{The neglected terms are of relative order $(m_e^2/s)\ln(s/m_e^2)$. At the two-pion threshold they change $K_e(s)$ by $1\times10^{-4}$ and, owing to the cancellation, $K_{\mu-e}(s)$ by $3\times10^{-4}$; even at the $\pi^0\gamma$ threshold the changes stay below $4\times10^{-4}$ and $7\times10^{-4}$, respectively, and they decrease with $s$. The approximate kernel is used in the analytic result \eref{auepi0gamma_ana}, whose value changes by only $0.02\%$ when the exact kernels are used.} ---  take the limit $m_e^2/s \to 0$ and obtain
\beq
a^{\mathrm{HVP}}_e 
\cong \frac{\alpha m_e^2}{3\pi}  \Pi'(0) \equiv 
\frac{\alpha}{3\pi^2} m_e^2
\int\limits_{s_0}^\infty \dd s \, \frac{\im \Pi(s)}{s^2} .
\eqlab{ae_slope}
\eeq 
This expression makes explicit that the high-energy contributions scale with the square of the lepton mass. Therefore, they are expected to cancel in $\aue$, introduced in \Eqref{aue_og_def}. Indeed, this quantity obeys an analogous representation, but with the kernel
\bea\label{eq:Kue}
K_{\mu-e}(s) &=&  K_\mu(s) - \frac{m_\mu^2}{m_e^2} K_e(s) \nn\\
&\cong &
-\frac{m_\mu^4}{s^2}
\int\limits_0^1 \dd x 
\frac{ x^4}
{1-x+x^2 (m_\mu^2/s)} \\
&=& K_\mu(s) - \frac{m_\mu^2}{3s} \, , \nn
\eea
which contains an additional power of $1/s$, thereby suppressing the high-energy region. In Fig.~\ref{fig:kernel_ratio},
we plot the ratio of $\mu-e$ over $\mu$ kernels to illustrate this point. 

The dispersive representation is, of course, mainly relevant to the data-driven approach, 
but analogous arguments can be made for representations employed in lattice QCD. Since the rescaled $\aehvp$ is essentially given by the slope of the vacuum polarization at zero momentum, \Eqref{ae_slope}, $\auehvp$ is the muon HVP with this leading Taylor term subtracted; both ingredients are routinely computed in lattice QCD, so that $\auehvp$ can be obtained from existing lattice data. For example, in the time-momentum representation of Bernecker and Meyer~\cite{Bernecker:2011gh}, the short-time contributions are suppressed; the leading 
$t^4$ behavior, at small $t$, is canceled and the expansion begins with $t^6 \log t$. This, in principle, makes the calculation of $\aue$ less prone to 
short-distance systematics (\eg, discretization errors \cite{Beltran:2026ofp}) than
that of $\au$ and $\ae$ separately. It will be interesting to see this realized in practice. On the other hand, the weight is shifted to long distances: for the standard windows of Ref.~\cite{RBC:2018dos} (boundaries at 0.4 and 1.0~fm, smoothed over 0.15~fm), the short-distance, intermediate, and long-distance parts of $\auehvp$ amount to about 1\%, 14\%, and 85\%, compared with 10\%, 33\%, and 57\% for $\auhvp$~\cite{Colangelo:2022vok}. (We estimate the former with a simple model of $R(s)$ that reproduces the data-driven values of $\auhvp$, $\aehvp$, and $\auehvp$, as well as the $\auhvp$ windows, to within 1--3\%.) A lattice determination of $\auehvp$ will therefore be limited by noise and finite-volume effects at long distances rather than by discretization effects.
\section{An illustration for $\pi^+\pi^-$ channel }
\label{sec:kernel_fucntion_method}
The largest part of the HVP contribution in the data-driven approach comes from the $\pi^+\pi^-$ channel. It is also one of the most controversial, where 
several ($e^+ e^- \to \pi^+\pi^-$) datasets have substantial disagreements \cite{Aliberti:2025beg}.

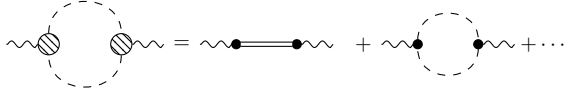
\begin{figure}[htb]
\begin{tikzpicture}[scale=0.8, transform shape]
\begin{feynman}
\vertex (a) at (-5.2,0);
\node[blob, minimum size=3.5mm] (b) at (-4.5,0);
\node[blob, minimum size=3.5mm] (c) at (-3.3,0);
\vertex (d) at (-2.6,0);
\node at (-2.3,0) {$=$};
\vertex (a1) at (-2,0);
\node[dot] (b1) at (-1.4,0);
\node[dot] (c1) at (-0.4,0);
\vertex (d1) at (0.2,0);
\node at (0.7,0) {$+$};
\vertex (a2) at (1,0);
\node[dot] (b2) at (1.6,0);
\node[dot] (c2) at (2.6,0);
\vertex (d2) at (3.2,0);
\node at (3.7,0) {$+\cdots$};
\diagram*{
  (a) -- [photon] (b) -- [scalar, half left, looseness=1.5] (c) -- [photon] (d),
  (c) -- [scalar, half left, looseness=1.5] (b),
  (a1) -- [photon] (b1) -- [doublefermion, edge label=\(\rho\)] (c1) -- [photon] (d1),
  (a2) -- [photon] (b2) -- [scalar, half left, edge label=\(\pi\)] (c2) -- [photon] (d2),
  (c2) -- [scalar, half left] (b2),
};
\end{feynman}
\end{tikzpicture}
    \caption{Two-pion contribution to the HVP in the VMD model. Blobs denote the pion electromagnetic form factor, the dashed line the pion propagator, and the double line the dressed $\rho$-meson propagator. }
    \label{fig:pipi_feynman_diagrams}
\end{figure}

On the theory side, this contribution can roughly be understood using a vector-meson-dominance (VMD) type model, where the photon couples to the pion via the 
$\rho$-meson. The corresponding HVP contribution is depicted in \Figref{pipi_feynman_diagrams}, leading to the well-known generic expression for the spectral function:
\begin{equation}\label{eq:ImPi_2pi}
    \im\Pi^{(\pi\pi)}(s)=
      \frac{\alpha}{12} \Big( 1- \frac{4 m^2_\pi}{s}\Big)^{\frac{3}{2}} 
      |F_\pi (s)|^2 \;\theta (s-4m_\pi^2) , 
\end{equation}
which is the one-loop scalar-QED expression \cite{Peskin:1995ev}
augmented with the pion electromagnetic form factor, $F_\pi(q^2)$. 
In the VMD picture, the latter is essentially described by the $\rho$-meson exchange, as seen in \Figref{pipi_feynman_diagrams}.
A more systematic description can be obtained in $\chi$PT supplemented by explicit vector-meson degrees of freedom, which would allow one to better estimate the theoretical uncertainties.
Our present analysis, however, is only illustrative.

\begin{table*}[htb]\begin{minipage}{\linewidth}
\renewcommand{\thempfootnote}{\alph{mpfootnote}} 
\renewcommand{\arraystretch}{2}
\begin{ruledtabular}
\begin{tabular}{c|lll}
 Source & $\au^{\pi^+\pi^-} (10^{-10})$  & $\ae^{\pi^+\pi^-} (10^{-14})$ & $\aue^{\pi^+\pi^-} (10^{-10})$\\
 \hline
VMD model \cite{Biloshytskyi:2025fjm}  &  $508.6$   & $140.0$ & $-90.0$\\
KNT eval.~\cite{Keshavarzi:2019abf} &  $503.46(1.91)$ & $138.59(54)$ & $-89.1(4)$\footnotemark[1]\\
\end{tabular}  
\end{ruledtabular}
\footnotetext[1]{Obtained from the previous two columns in the strong-correlation limit.}
\caption{The $\pi^+\pi^-$-channel contribution in a VMD model compared with a state-of-the-art data-driven evaluation.
The VMD results serve a qualitative purpose and no effort to estimate uncertainties is made here.
}
  \label{tab:pipi_breakdown}
    \end{minipage}
\end{table*}

\begin{figure}[htb]
    \centering    \includegraphics[width=0.9\linewidth]{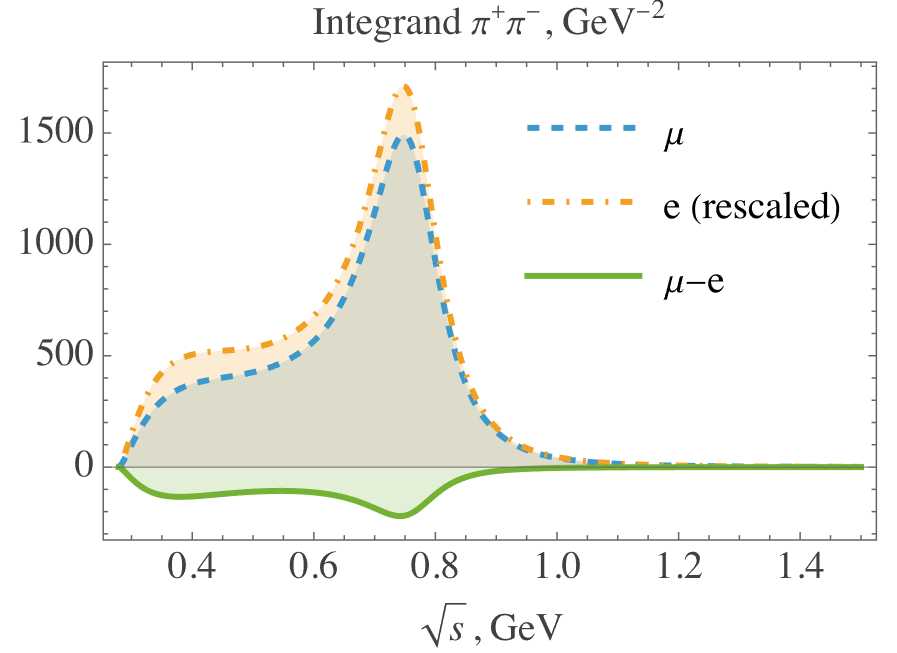}
    \caption{
    The integrand of $a_\ell$, Eq.~\eqref{eq:HVPdisp_og}, in the $\pi^+\pi^-$ channel (in units of $10^{-10}$) for the muon, the rescaled electron, and their difference $\mu-e$, obtained with the kernel functions~\eqref{eq:Ku} and \eqref{eq:Kue} and the VMD spectral function~\eqref{eq:ImPi_2pi}.   
    }
    \label{fig:integrand_plot}
\end{figure}

Using the specific VMD model of Ref.~\cite{Biloshytskyi:2025fjm}, we obtain the values in Table~\ref{tab:pipi_breakdown}, where they
are compared to a state-of-the-art data-driven evaluation~\cite{Keshavarzi:2019abf}. Fig.~\ref{fig:integrand_plot} shows
the corresponding integrands as functions of $\sqrt{s}$, obtained with Eqs.~\eqref{eq:Ku}, \eqref{eq:Kue}, and \eqref{eq:ImPi_2pi}, illustrating the large cancellation between
the muon and rescaled electron $g-2$. Sub-GeV contributions, such as the $\rho$-meson resonance, survive in $\aue$, while higher-energy contributions are suppressed by an additional power of $m_\mu^2/s$.

The table shows a qualitative agreement between the model 
and the data-driven evaluation, at the 1\% level for all three quantities; in absolute terms, the difference is much smaller for $\aue$ than for $\au$.
More importantly, the combined quantity allows one to use the parameter-free $\chi$PT predictions for contributions that are otherwise model dependent. In the following section we illustrate this for an isospin-breaking correction, the $\pi^0\gamma$ contribution.

\section{Pseudoscalar-meson contribution in  $\chi$PT}
\seclab{pi0gamma}

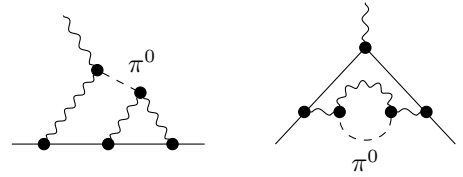
\begin{figure}[bth]
\centering
\begin{tikzpicture}[scale=0.9]
\begin{feynman}

\vertex (a) at (-3.2,1.0){\(\gamma\)};
\node[dot, minimum size=1mm] (b) at (-2.67,0.15);
\node[dot, minimum size=1mm] (c) at (-2,-0.2);
\vertex (d) at (-4,-1){\(\ell\)};
\node[dot, minimum size=1mm] (e) at (-3.5,-1);
\node[dot, minimum size=1mm] (f) at (-2.5,-1);
\node[dot, minimum size=1mm] (g) at (-1.5,-1);
\vertex (h) at (-1,-1){\(\ell\)};

\node[dot, minimum size=1mm] (a2) at (0.55,-0.5);
\node[dot, minimum size=1mm] (b2) at (1.1,-0.5);
\node[dot, minimum size=1mm] (c2) at (1.9,-0.5);
\node[dot, minimum size=1mm] (d2) at (2.45,-0.5);
\node[dot, minimum size=1mm] (e2) at (1.5,0.5);
\vertex (f2) at (0.1,-1){\(\ell\)};
\vertex (g2) at (2.9,-1){\(\ell\)};
\vertex (h2) at (1.5,1.2){\(\gamma\)};
\diagram*{
  (a) -- [photon] (b)
      -- [scalar, edge label=\(\pi^0\)] (c)
      -- [photon] (f),
  (c) -- [photon] (g),
  (b) -- [photon] (e),
  (d) -- [plain] (h),

  (a2) -- [photon] (b2)
       -- [photon, half left] (c2)
       -- [photon] (d2),
  (c2) -- [scalar, half left, edge label=\(\pi^0\)] (b2),
  (h2) -- [photon] (e2),
  (f2) -- [plain] (e2) -- [plain] (g2),
};

\end{feynman}
\end{tikzpicture}
    \caption{Pseudoscalar-meson contribution to light-by-light (left) and isospin-breaking HVP (right).}
    \label{fig:lbl_pi0gamma_feynman_diagrams}
\end{figure}

The pseudoscalars $(\pi^0, \eta, \eta')$ dominate
the hadronic light-by-light contribution (HLbL), and also
prominently appear in the isospin-breaking correction to the HVP
contribution. The corresponding diagrams are shown in \Figref{lbl_pi0gamma_feynman_diagrams}, where we focus on $\pi^0$; the other pseudoscalars (including ALPs) can be treated analogously.

One problem with these
diagrams is that the $\pi\gamma\gamma$ coupling derives from
the well-known Wess-Zumino-Witten (WZW) Lagrangian, which contains
a dimension-5 operator,
\beq
{\cal L} = -\frac{\al}{4\pi f_\pi} \pi^0 F_{\mu\nu} \tilde F^{\mu\nu},
\eeq
with $\pi^0 $ the pseudoscalar field, and $F$, $\tilde F$ the electromagnetic field strength and its dual; $f_\pi \simeq 92.4$~MeV the pion-decay constant.  This results in an ultraviolet (UV) divergent contribution to $g-2$. The divergence can be ameliorated by inserting an empirical pion transition form factor (TFF) in the $\pi\gamma\gamma$ vertex\footnote{In the ALP case, the divergence is renormalized by a local counterterm (Wilson coefficient), rendering the prediction dependent on an unknown low-energy constant, which is then omitted in a given
renormalization scheme \cite{Bauer:2020jbp,Bauer:2017nlg,Bauer:2017ris}. The same procedure could in principle be implemented for the pion contributions in $\chi$PT; however, it obviously has less predictive power than using an empirical TFF.}, but this opens another set of issues, such as 
the implementation of short-distance constraints \cite{Melnikov:2003xd,Bijnens:2019ghy,Colangelo:2019lpu}.

Fortunately, this UV-divergence exactly cancels  from the combined quantity $a_{\mu-e}$, which allows us to consider the $\chi$PT predictions as they are.
Explicitly, with a UV cutoff $\Lambda$, the pointlike vertex gives
\beq
a_\ell^{\pi^0\gamma} = \frac{\alpha^3 m_\ell^2}{288\pi^5 f_\pi^2}\ln\frac{\Lambda^2}{m_{\pi^0}^2} + \dots ,
\eeq
with the ellipsis denoting finite terms. This is of the form \eref{LECgeneric}: in $\chi$PT, the logarithm is renormalized by the flavor-universal lepton-dipole counterterm~\cite{Knecht:2001qg,Ramsey-Musolf:2002gmi}, and both drop out of $\aue$.
Here we only consider the $\pi^0 \gamma$ contribution via HVP, see \Figref{lbl_pi0gamma_feynman_diagrams} (right). 

Substituting the corresponding expression for the spectral function, \ie,
\bea
        \im \Pi^{(\pi^0\gamma)}(s) &=& 
         \frac{\pi\alpha^2}{6} \frac{s}{\ga4\pi^2 f_\pi \dr^2} \nn\\
         &&\times \Big( 1-\frac{m_{\pi^0}^2}{s} \Big)^3 \theta\ga s - m_{\pi^0}^2\dr,
\eea
into the dispersive representation \eref{HVPdisp_og}, with the kernel \eref{Kue}, we obtain (for $r = m_\mu/m_{\pi^0}$):
\begin{subequations}
    \bea
        a_{\mu-e}^{\pi^0\gamma} & = &\frac{\alpha^3 m_{\pi^0}^2}{1728\,\pi^5 \, f_\pi^2\, r^4 }
    \Big[ r^2 \big(  4r^4+81r^2+66 \big) \nonumber\\
       && - \, \frac{6 \big( 64 r^4 + 28 r^2 -11 \big)}{\sqrt{4 r^2-1}} \arccos{\frac{1}{2r}}
        \nonumber\\  
       && - \, 6 (2r^6+9r^4-11)\log r \nonumber\\
       && + \, 18(6r^2+1) \Big(\log^2 r + \arccos^2{\frac{1}{2r}}\Big)
        \Big]\eqlab{auepi0gamma_ana}\\
        &= &  -0.34 (5) \times 10^{-11}\,, \eqlab{auepi0gamma}
\eea
\end{subequations}
where in the resulting numerical value we include about 15\% uncertainty representing an estimate of higher-order contributions.

This value can be compared to the contribution to the muon alone~\cite{Blokland:2001pb,Biloshytskyi:2025fjm}:
\beq
a_{\mu}^{\pi^0\gamma} \simeq 1 \times 10^{-11} .
\eeq
The latter quantity depends on some form of regulator, for example, the empirical pion transition form factor. Performing the same calculation for the electron (with the TFF of Ref.~\cite{Biloshytskyi:2025fjm}) and forming the combination, we obtain
\beq
a_{\mu-e}^{\pi^0\gamma} \mbox{(with TFF)} \simeq - 0.45 \times 10^{-11},\eqlab{amupi0gamma_TFF}
\eeq
which is comparable to the model-independent result \eref{auepi0gamma}, but far from being in perfect agreement.\footnote{These are field-theoretic results for the $\pi^0\gamma$ contribution. Data-driven evaluations of the $\pi^0\gamma$ channel, based on the measured $e^+e^-\to\pi^0\gamma$ cross section, give much larger values, \eg, $\aue^{\pi^0\gamma}\approx -6.8\times10^{-11}$ from the values of Ref.~\cite{Keshavarzi:2019abf}; the origin of this difference is explained in Ref.~\cite{Biloshytskyi:2025fjm}.}

It will be interesting to perform a similar comparison for the HLbL contribution, \Figref{lbl_pi0gamma_feynman_diagrams} (left), and see which of the
model-dependent predictions agrees better with $\chi$PT. 
 The pointlike WZW
vertices make this two-loop contribution UV-divergent, with the leading behavior given by~\cite{Knecht:2001qg,Ramsey-Musolf:2002gmi}
\beq\eqlab{aell_lbl}
a_\ell^{\pi^0,\,\mathrm{LbL}} = \ga\frac{\al}{\pi}\dr^{3}
\frac{3 m_\ell^{2}}{16\pi^{2} f_\pi^{2}}\,
\ln^{2}\!\frac{\Lambda}{m_{\pi^0}} + \dots,
\eeq
where $\Lambda$ is the UV cutoff and 
the ellipsis stands for single-logarithmic and finite terms. The coefficient of
the double logarithm is universal --- both leptons being lighter than the pion, it
multiplies $m_\ell^{2}$ by a function of $\Lambda/m_{\pi^0}$ alone \cite{Marciano:2016yhf} --- hence, cancels in $\aue$.  A detailed two-loop evaluation is deferred to a separate publication.

\section{Ultimate sensitivity to New Physics}
\label{sec:newphysics}

One of the reasons the $g-2$ of the muon attracted so much attention is its potential sensitivity to the BSM effects from, {\eg},  loop diagrams with exchange of new particles.
Any test of New Physics is limited by the experimental precision of the measurements and by the uncertainty in $a_\mu^{\rm HVP}$.
In this section, we argue that $a_{\mu-e}$ provides a deeper reach into the parameter space of New Physics compared with $a_\mu$ or $a_e$ considered separately. Such a test, however, has additional limitations. In particular, {\em flavor-universal} New Physics, parametrized by a dimension-5 contact operator of the same form as the counterterm in \Eqref{Lct}, but with a New-Physics coefficient, gives corrections to $a_\ell$ proportional to the square of the lepton mass, $\Delta a_\ell\propto m_\ell^2 \Lambda^{-2}$, and therefore cancels out of $a_{\mu-e}$.

Flavor universality, however, does not have to hold exactly. In supersymmetric models, for example, the masses of first and second generations of sleptons (superpartners of electrons and muons) do not have to be equal, resulting in a different value of loop integrals, which translates to two different effective values of $\Lambda$ for the muon and electron.  In this case, the New Physics correction to $a_{\mu-e}$ can be written as
\begin{equation}
\label{nonuniv}
     a^{\rm BSM}_{\mu-e}= m_\mu^2 \left(
    \frac{1}{\Lambda^2_{\mu}}- \frac{1}{\Lambda^2_{e}}\right).
\end{equation}
A new gauge symmetry based on anomaly-free combinations of the Standard Model charges, such as the difference between specific flavors, \eg, $L_\mu-L_\tau$, will also provide non-universal contributions to $a_\ell$. 
It is clear from the preceding discussion that the ultimate sensitivity to $\Lambda^{-2}_{\mu}- \Lambda^{-2}_{e}$ in Eq.\,(\ref{nonuniv}) is better than to $\Lambda^{-2}_{\mu}$ and $\Lambda^{-2}_{e}$ separately; if the error is dominated by HVP, the improvement is given by the factor $\delta a_\mu^{\rm HVP}/\delta a_{\mu-e}^{\rm HVP}\approx 6$ (cf.\ the data-driven entries of Table~\ref{tab:ref_values}).

It is worth stressing what separates this from an optimized window on $\au$ alone. The flavor-universal contribution, being $\propto m_\ell^{2}$, is present in $\au$ in full and cannot be removed by any choice of weight function; it cancels only against the rescaled $\ae$, that is, only in the physical combination $\aue$. Reweighting $\au$ can sharpen the sensitivity to a given energy region, but it leaves the universal piece in place and therefore cannot, by itself, isolate the non-universal signal of Eq.\,(\ref{nonuniv}). The gain in reach shown below is thus a property of the observable, not of the weight, and is not reproducible by window optimization within $\au$.

Next we address the universally coupled New Physics, but with the mass scale much below the weak scale. As an example, we take the so-called dark photon that corrects $a_{\mu(e)}$ at one loop \cite{Fayet:2007ua,Pospelov:2008zw,Arkani-Hamed:2008kxc}. The model is very simple and UV-complete. Its low-energy limit contains only two extra parameters: 
\begin{eqnarray}
    {\cal L}_{d.\,ph.} &=&-\frac14 (F_{\mu\nu})^2 -\frac14 (F_{\mu\nu}')^2 + \frac12 m_{A'}^2 A_\mu^{\prime\,2}\nonumber \\ 
   & +& \sum_\ell \bar \psi_\ell \big[\gamma^\mu(i\partial_\mu - e A_\mu - e\varepsilon A_\mu')-m_\ell\big]\psi_\ell,
\label{DPh}
\end{eqnarray}
where $A_\mu$ and $A'_\mu$ are the photon and dark-photon fields with field strengths $F_{\mu\nu}$ and $F'_{\mu\nu}$, $m_{A'}$ is the dark-photon mass, and $\varepsilon$ is the kinetic-mixing parameter.
Corrections to the leptonic $a_{\mu(e)}$ can be written as \cite{Pospelov:2008zw}
\begin{eqnarray}
    a_\ell^{A'} = \frac{\alpha}{2\pi}\times \varepsilon^2 \int_0^1 \dd x \frac{2 m_\ell^2x^2(1-x)}{m_\ell^2x^2+m_{A'}^2(1-x)}.
    \label{aDP} 
\end{eqnarray}
We note in passing that it is of course equivalent to using a new delta-functional contribution in the polarization operator, $\im \Pi(s)\propto \varepsilon^2 \delta(s-m_{A'}^2)$, inside the dispersive formula.

In the limit of $m_{A'} \gg m_\mu $, Eq.\,(\ref{aDP}) leads to the universal pattern of anomalous magnetic moments, $a_\ell \propto m_\ell^{2}m_{A'}^{-2}$, to which $a_{\mu-e}$ is insensitive. However, the most interesting part of the mass range corresponds to sub-GeV dark photons \cite{Fayet:2007ua,Pospelov:2008zw}, where the $g-2$ measurements can be competitive with direct searches. 

\begin{figure}
    \centering
    \includegraphics[width=\linewidth]{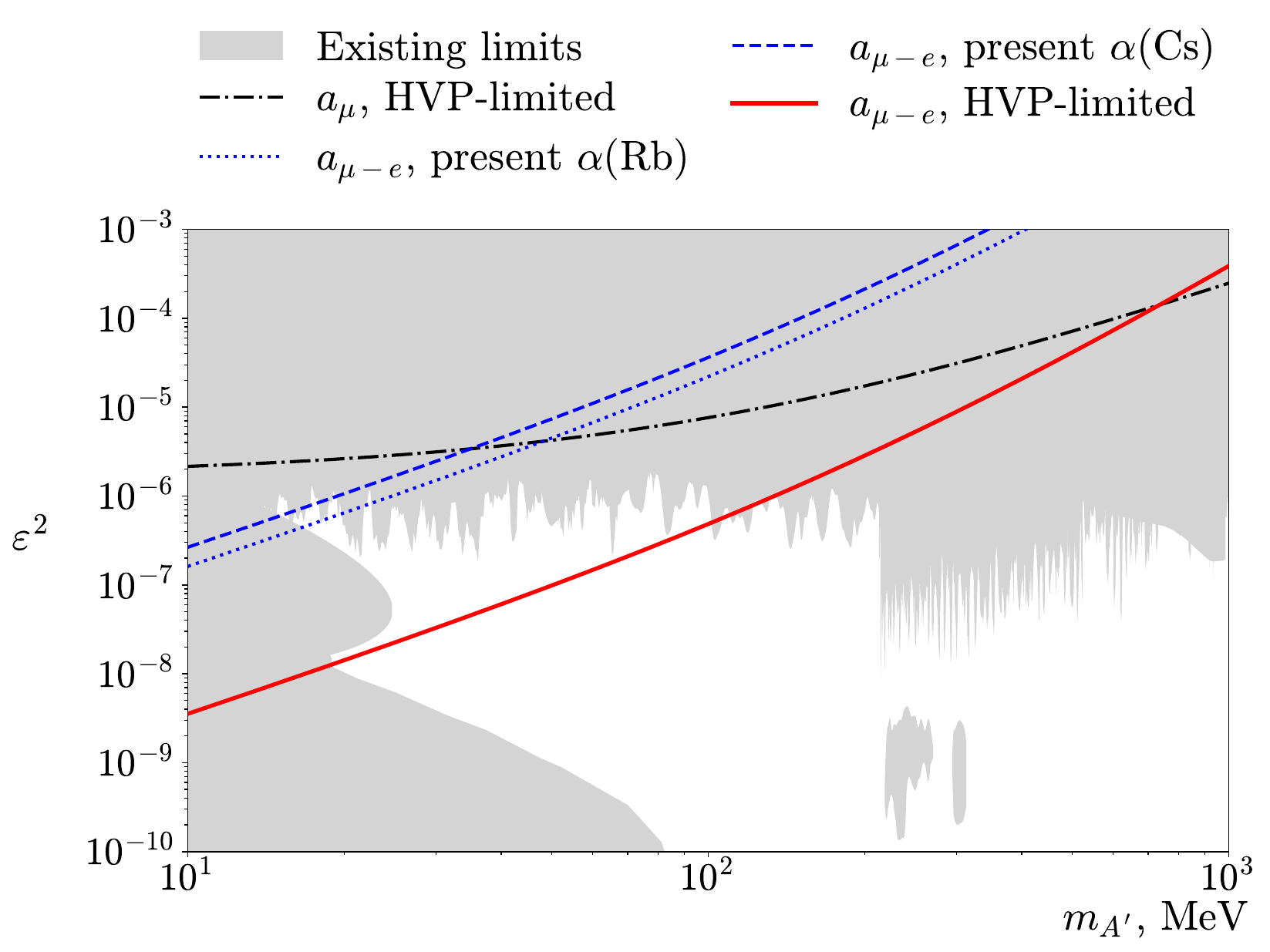}
    \caption{Exclusion plot for the dark-photon parameters; the parameter space above the curves would be excluded at 95\% CL. The dash-dotted line assumes that the only limitation on $\au$ is an HVP error $\delta a_\mu^{\rm HVP}=10^{-9}$, and the red line likewise assumes $\delta a_{\mu-e}^{\rm HVP}=0.15\,\delta a_\mu^{\rm HVP}$. The blue lines show the reach of $\aue$ with the present uncertainties of $a_e$~\cite{Fan:2022eto} and of $\alpha$ from Cs~\cite{Parker:2018vye} and Rb~\cite{Morel:2020dww} measurements. Gray: existing limits (mostly at 90\% CL) as compiled in DarkCast~\cite{Ilten:2018crw}, supplemented by FASER~\cite{FASER:2023tle}.
    }
\label{fig:Figure_DPh}
\end{figure}

In order to demonstrate the ultimate constraining power of $a_{\mu-e}$ we are going to assume the following hypothetical ({\em i.e.} distant future) scenario: {\em i.} Experiments measuring $a_\mu$, $a_e$, and $\alpha$ become so precise that the experimental errors are negligible, {\em ii.} Likewise, non-QCD errors (QED, EW contributions) are calculated to sufficiently high order not to matter either, {\em iii.} Constraining power of the $g-2$ experiments is still limited by the hadronic contributions, and HVP specifically. In other words, we are assuming that the error is entirely controlled by $\delta a_\mu^{\rm HVP}$ and $\delta a_e^{\rm HVP}\simeq (m_e/m_\mu)^2 \delta a_\mu^{\rm HVP}$. These errors are, of course, highly correlated, and accounting for their correlation yields a hadronic error for $a_{\mu -e}$ about a factor of 6 smaller than $\delta a_\mu^{\rm HVP}$ (cf.\ the data-driven entries of Table~\ref{tab:ref_values}); for illustration we take $\delta a_{\mu-e}^{\rm HVP}\simeq 0.15 \, \delta a_{\mu}^{\rm HVP}$.
In that hypothetical regime, the constraints (or more precisely, sensitivity limits) on the dark photon can be substantially improved, see Fig.\,\ref{fig:Figure_DPh}. For $m_{A'}\lesssim 740$~MeV, $\aue$ would provide a deeper reach into $\varepsilon^2$ than $\au$, by factors of about 600, 100, and 16 at $m_{A'}=10$, 30, and 100~MeV, respectively. At $m_{A'}\simeq 740$~MeV, $|a^{A'}_{\mu-e}|/a^{A'}_\mu$ drops below $\delta a_{\mu-e}^{\rm HVP}/\delta a_\mu^{\rm HVP}$ and the gain disappears. Existing direct searches for visible dark-photon decays already exclude the corresponding couplings for $m_{A'}\lesssim 20$~MeV and $m_{A'}\gtrsim 100$~MeV, leaving a mass range of about 20--100~MeV in which $\aue$ would probe new parameter space.

This figure shows that the sub-GeV dark photons could be better probed by $\aue$, conditional on experimental improvement of $a_e$ and $\alpha$ (as well as some theoretical improvement of the high-order QED contributions)\,\cite{Gabrielse:2025jep}. For $m_{A'}\gtrsim 740$~MeV, the flavor-universal contribution dominates,  $a_\ell^{A'}\propto m_\ell^2 m_{A'}^{-2}$, and $\aue$ is not adding sensitivity. 
An equivalent restatement of possible improvements using $a_{\mu-e}$ can be formulated like this: one could extract the main part of the hadronic corrections from the measurement of $a_\mu$, and carry it over into the prediction of $a_e$, with small theoretical corrections calculated explicitly (mainly related to $a^{\rm HVP}_{\mu-e}$). Then the comparison of experimental results for $a_e$ and $\alpha$ with ``$a_\mu$-informed'' theoretical value of $a_e$ would provide a stringent test of the Standard Model. 

The same sensitivity curves apply to other light vector bosons whose couplings to electrons and muons have equal magnitude, such as the $B-L$ and $L_e-L_\mu$ gauge bosons, with $e\varepsilon$ replaced by the gauge coupling.

\section{Conclusion and Outlook}
Window quantities have become a standard tool for confronting lattice and data-driven results. We have introduced a sub-GeV window quantity of a different kind, $\aue$, constructed by exploiting the strong correlation of the hadronic vacuum polarization (HVP) contributions to the anomalous magnetic moments of the muon and the electron, in particular at high energies. More generally, assuming that the high-energy effects scale with $m_\ell^2$, they all cancel from $\aue$. This is analogous to hydrogen-like systems, where short-range effects, scaling with the third power of the reduced mass, can be arranged to cancel between, \eg, muonic and ordinary hydrogen.

The subtraction of the rescaled $\ae$ from $\au$ can be reformulated as a modification of the kernel function, which acts as a soft cutoff of about 1~GeV. Given the ongoing tension between the data-driven and lattice-QCD determinations of the HVP contribution to $\au$, the large cancellation of uncertainties in $\auehvp$ presents a clear advantage.

The examples worked out in this paper illustrate these features. In the $\pi^+\pi^-$ channel, most of the muon contribution is canceled by the rescaled electron contribution; the remainder comes from the $\rho$-meson region, where even a simple vector-meson-dominance model lies close to the data-driven value. For the $\pi^0\gamma$ contribution, the cancellation removes the unknown low-energy constant, so that chiral perturbation theory gives a parameter-free prediction. For a dark photon, the smaller hadronic uncertainty translates into a reach in the kinetic mixing well beyond that of $\au$ alone, including masses between 20 and 100~MeV that are not yet excluded by direct searches.

While hadronic uncertainties are significantly suppressed, the precision of $\aue$ becomes limited by the uncertainties in $\ae$ and in the fine-structure constant. Since the SM prediction for $\ae$ relies on $\alpha$ as an external input, the uncertainty in $\alpha$, like that of the measured $\ae$, enters $\aue$ enhanced by $(m_\mu/m_e)^2$, and the present tension between the Cs and Rb determinations of $\alpha$ has to be resolved first. Future improvements in $\alpha$ and $\ae$, to $\delta\alpha/\alpha\lesssim 2\times10^{-12}$ and $\delta a_e\lesssim 2\times10^{-15}$, would make $\aue$ a complementary probe of low-energy hadronic effects and a benchmark for both dispersive and lattice-QCD evaluations.

Unlike the Euclidean windows used to compare lattice and dispersive evaluations, $\aue$ is an experimental observable in its own right, and the exact cancellations it enforces (of the chiral low-energy constant and of flavor-universal New Physics) are beyond the reach of any reweighting of $\au$. On the theory side, the natural next steps are a direct lattice computation of $\auehvp$, a data-driven evaluation that keeps the full correlation between the muon and electron contributions, and a two-loop chiral calculation of the pseudoscalar contribution to hadronic light-by-light scattering. Together with better measurements of $\ae$ and $\alpha$, these would make $\aue$ a test of the Standard Model in which the hadronic uncertainty is no longer the limiting factor.

\section*{Acknowledgements}
M.~P. thanks Gerald~Gabrielse for the initial discussion of hadronic uncertainties in $a_e$ that led to this study; S.~L. thanks Volodymyr Biloshytskyi for helpful discussions, while V.~P.\  acknowledges insightful conversations with Bogdan Malaescu, Harvey Meyer, and Matthias Neubert.

This work is supported by the Deutsche Forschungsgemeinschaft (DFG) 
through the Collaborative Research Center 1660 ``Hadrons and Nuclei as Discovery Tools'' (grant 514321794). M.~P.'s work is supported in part by the DOE grant DE-SC0011842. M.~P. also
acknowledges the financial support provided by CERN.

In preparing the revised version, the authors used Claude Opus~5.5 (Anthropic) to cross-check calculations, write the plotting code for Fig.~\ref{fig:Figure_DPh}, and draft parts of the revised text; all output was checked by the authors, who take full responsibility for the content.

\bibliography{references.bib}

@article{Hansen:2019idp,
    author = "Hansen, Martin and Lupo, Alessandro and Tantalo, Nazario",
    title = "{On the extraction of spectral densities from lattice correlators}",
    eprint = "1903.06476",
    archivePrefix = "arXiv",
    primaryClass = "hep-lat",
    doi = "10.1103/PhysRevD.99.094508",
    journal = "Phys. Rev. D",
    volume = "99",
    number = "9",
    pages = "094508",
    year = "2019"
}

@article{Giudice:2012pf,
    author = "Giudice, Gian F. and Paradisi, Paride and Strumia, Alessandro ",
    title = "{Correlation between the Higgs Decay Rate to Two Photons and the Muon g - 2}",
    eprint = "1207.6393",
    archivePrefix = "arXiv",
    primaryClass = "hep-ph",
    reportNumber = "CERN-PH-TH-2012-209",
    doi = "10.1007/JHEP10(2012)186",
    journal = "JHEP",
    volume = "2012",
    number = "10",
    pages = "186",
    year = "2012"
}

@article{DiLuzio:2024sps,
    author = "Di Luzio, Luca and Keshavarzi, Alexander and Masiero, Antonio and Paradisi, Paride",
    title = "{Model-Independent Tests of the Hadronic Vacuum Polarization Contribution to the Muon g-2}",
    eprint = "2408.01123",
    archivePrefix = "arXiv",
    primaryClass = "hep-ph",
    doi = "10.1103/PhysRevLett.134.011902",
    journal = "Phys. Rev. Lett.",
    volume = "134",
    number = "1",
    pages = "011902",
    year = "2025"
}

@misc{Gabrielse:2025jep,
    author = "Gabrielse, Gerald and Venanzoni, Graziano",
    title = "{Measured Lepton Magnetic Moments}",
    eprint = "2507.11268",
    archivePrefix = "arXiv",
    primaryClass = "hep-ex",
    month = "7",
    year = "2025"
}

@article{Arkani-Hamed:2008kxc,
    author = "Arkani-Hamed, Nima and Weiner, Neal",
    title = "{LHC Signals for a SuperUnified Theory of Dark Matter}",
    eprint = "0810.0714",
    archivePrefix = "arXiv",
    primaryClass = "hep-ph",
    doi = "10.1088/1126-6708/2008/12/104",
    journal = "JHEP",
    volume = "2008",
    number = "12",
    pages = "104",
    year = "2008"
}

@article{Fayet:2007ua,
    author = "Fayet, Pierre",
    title = "{U-boson production in e+ e- annihilations, psi and Upsilon decays, and Light Dark Matter}",
    eprint = "hep-ph/0702176",
    archivePrefix = "arXiv",
    reportNumber = "LPTENS-07-07",
    doi = "10.1103/PhysRevD.75.115017",
    journal = "Phys. Rev. D",
    volume = "75",
    pages = "115017",
    year = "2007"
}

@article{Pospelov:2008zw,
    author = "Pospelov, Maxim",
    title = "{Secluded U(1) below the weak scale}",
    eprint = "0811.1030",
    archivePrefix = "arXiv",
    primaryClass = "hep-ph",
    doi = "10.1103/PhysRevD.80.095002",
    journal = "Phys. Rev. D",
    volume = "80",
    pages = "095002",
    year = "2009"
}

@article{Bernecker:2011gh,
    author = "Bernecker, David and Meyer, Harvey B.",
    title = "{Vector Correlators in Lattice QCD: Methods and applications}",
    eprint = "1107.4388",
    archivePrefix = "arXiv",
    primaryClass = "hep-lat",
    doi = "10.1140/epja/i2011-11148-6",
    journal = "Eur. Phys. J. A",
    volume = "47",
    pages = "148",
    year = "2011"
}

@article{Aoyama:2020ynm,
	archiveprefix = {arXiv},
	author = {Aoyama, T. and others},
	doi = {10.1016/j.physrep.2020.07.006},
	eprint = {2006.04822},
	journal = {Phys. Rept.},
	pages = {1--166},
	primaryclass = {hep-ph},
	reportnumber = {FERMILAB-PUB-20-207-T, INT-PUB-20-021, KEK Preprint 2020-5, MITP/20-028, KEK Preprint 2020-5, MITP/20-028, CERN-TH-2020-075, IFT-UAM/CSIC-20-74, LMU-ASC 18/20, LTH 1234, LU TP 20-20, LTH 1234, LU TP 20-20, MAN/HEP/2020/003, PSI-PR-20-06, UWThPh 2020-14, ZU-TH 18/20},
	title = {{The anomalous magnetic moment of the muon in the Standard Model}},
	volume = {887},
	year = {2020}}

@article{Colangelo:2022vok,
	archiveprefix = {arXiv},
	author = {Colangelo, G. and El-Khadra, A. X. and Hoferichter, M. and Keshavarzi, A. and Lehner, C. and Stoffer, P. and Teubner, T.},
	doi = {10.1016/j.physletb.2022.137313},
	eprint = {2205.12963},
	journal = {Phys. Lett. B},
	pages = {137313},
	primaryclass = {hep-ph},
	reportnumber = {FERMILAB-PUB-22-443-V, LTH 1304, PSI-PR-22-13, ZU-TH 20/22, FERMILAB-PUB-22-443-V},
	title = {{Data-driven evaluations of Euclidean windows to scrutinize hadronic vacuum polarization}},
	volume = {833},
	year = {2022}}

@article{Bauer:2020jbp,
    author = "Bauer, Martin and Neubert, Matthias and Renner, Sophie and Schnubel, Marvin and Thamm, Andrea",
    title = "{The Low-Energy Effective Theory of Axions and ALPs}",
    eprint = "2012.12272",
    archivePrefix = "arXiv",
    primaryClass = "hep-ph",
    reportNumber = "IPPP/20/69, MITP/20-070 SISSA 30/2020/FISI, ZH-TH-47/20",
    doi = "10.1007/JHEP04(2021)063",
    journal = "JHEP",
    volume = "2021",
    number = "04",
    pages = "063",
    year = "2021"
}

@article{Ramsey-Musolf:2002gmi,
    author = "Ramsey-Musolf, M. J. and Wise, Mark B.",
    title = "{Hadronic light by light contribution to muon g-2 in chiral perturbation theory}",
    eprint = "hep-ph/0201297",
    archivePrefix = "arXiv",
    reportNumber = "CALT-68-2372, MAP-282",
    doi = "10.1103/PhysRevLett.89.041601",
    journal = "Phys. Rev. Lett.",
    volume = "89",
    pages = "041601",
    year = "2002"
}

@article{Marciano:2016yhf,
    author = "Marciano, W. J. and Masiero, A. and Paradisi, P. and Passera, M.",
    title = "{Contributions of axionlike particles to lepton dipole moments}",
    eprint = "1607.01022",
    archivePrefix = "arXiv",
    primaryClass = "hep-ph",
    doi = "10.1103/PhysRevD.94.115033",
    journal = "Phys. Rev. D",
    volume = "94",
    number = "11",
    pages = "115033",
    year = "2016"
}

@article{Bauer:2017ris,
    author = "Bauer, Martin and Neubert, Matthias and Thamm, Andrea",
    title = "{Collider Probes of Axion-Like Particles}",
    eprint = "1708.00443",
    archivePrefix = "arXiv",
    primaryClass = "hep-ph",
    reportNumber = "MITP-17-047",
    doi = "10.1007/JHEP12(2017)044",
    journal = "JHEP",
    volume = "2017",
    number = "12",
    pages = "044",
    year = "2017"
}

@article{Bauer:2017nlg,
    author = "Bauer, Martin and Neubert, Matthias and Thamm, Andrea",
    title = "{LHC as an Axion Factory: Probing an Axion Explanation for $(g-2)_\mu$ with Exotic Higgs Decays}",
    eprint = "1704.08207",
    archivePrefix = "arXiv",
    primaryClass = "hep-ph",
    doi = "10.1103/PhysRevLett.119.031802",
    journal = "Phys. Rev. Lett.",
    volume = "119",
    number = "3",
    pages = "031802",
    year = "2017"
}

@article{Giusti:2020efo,
    author = "Giusti, D. and Simula, S.",
    title = "{Ratios of the hadronic contributions to the lepton $g-2$ from Lattice QCD+QED simulations}",
    eprint = "2003.12086",
    archivePrefix = "arXiv",
    primaryClass = "hep-lat",
    doi = "10.1103/PhysRevD.102.054503",
    journal = "Phys. Rev. D",
    volume = "102",
    number = "5",
    pages = "054503",
    year = "2020"
}

@article{Bauer:2019gfk,
    author = "Bauer, Martin and Neubert, Matthias and Renner, Sophie and Schnubel, Marvin and Thamm, Andrea",
    title = "{Axionlike Particles, Lepton-Flavor Violation, and a New Explanation of $a_\mu$ and $a_e$}",
    eprint = "1908.00008",
    archivePrefix = "arXiv",
    primaryClass = "hep-ph",
    reportNumber = "CERN-TH-2019-124, IPPP/19/64, MITP/19-053",
    doi = "10.1103/PhysRevLett.124.211803",
    journal = "Phys. Rev. Lett.",
    volume = "124",
    number = "21",
    pages = "211803",
    year = "2020"
}

@article{Boccaletti:2024guq,
    author = "Boccaletti, A. and others",
    title = "{Hybrid calculation of hadronic vacuum polarization in muon g {\ensuremath{-}} 2 to 0.48{\%}}",
    eprint = "2407.10913",
    archivePrefix = "arXiv",
    primaryClass = "hep-lat",
    doi = "10.1038/s41586-026-10449-z",
    journal = "Nature",
    volume = "653",
    number = "8114",
    pages = "373--377",
    year = "2026"
}

@misc{Beltran:2026ofp,
    author = "Beltran, Arnau and Conigli, Alessandro and Kuberski, Simon and Meyer, Harvey B. and Ottnad, Konstantin and Wittig, Hartmut",
    title = "{Higher-order hadronic vacuum polarization contribution to the muon $g-2$ from lattice QCD}",
    eprint = "2603.06806",
    archivePrefix = "arXiv",
    primaryClass = "hep-lat",
    reportNumber = "MITP-26-006, CERN-TH-2026-035",
    month = "3",
    year = "2026"
}

@article{RBC:2018dos,
    author = {Blum, T. and Boyle, P. A. and G{\"u}lpers, V. and Izubuchi, T. and Jin, L. and Jung, C. and J{\"u}ttner, A. and Lehner, C. and Portelli, A. and Tsang, J. T.},
    collaboration = "RBC, UKQCD",
    title = "{Calculation of the hadronic vacuum polarization contribution to the muon anomalous magnetic moment}",
    eprint = "1801.07224",
    archivePrefix = "arXiv",
    primaryClass = "hep-lat",
    doi = "10.1103/PhysRevLett.121.022003",
    journal = "Phys. Rev. Lett.",
    volume = "121",
    number = "2",
    pages = "022003",
    year = "2018"
}

@article{Karshenboim:2021jsc,
    author = "Karshenboim, Savely G. and Shelyuto, Valery A.",
    title = "{Hadronic vacuum-polarization contribution to various QED observables}",
    doi = "10.1140/epjd/s10053-021-00052-4",
    journal = "Eur. Phys. J. D",
    volume = "75",
    number = "2",
    pages = "49",
    year = "2021"
}

@article{Colangelo:2019lpu,
    author = "Colangelo, Gilberto and Hagelstein, Franziska and Hoferichter, Martin and Laub, Laetitia and Stoffer, Peter",
    title = "{Short-distance constraints on hadronic light-by-light scattering in the anomalous magnetic moment of the muon}",
    eprint = "1910.11881",
    archivePrefix = "arXiv",
    primaryClass = "hep-ph",
    reportNumber = "INT-PUB-19-050",
    doi = "10.1103/PhysRevD.101.051501",
    journal = "Phys. Rev. D",
    volume = "101",
    number = "5",
    pages = "051501",
    year = "2020"
}

@article{Galda:2023qjx,
    author = "Galda, Anne Mareike and Neubert, Matthias",
    title = "{ALP-LEFT Interference and the Muon $(g-2)$}",
    eprint = "2308.01338",
    archivePrefix = "arXiv",
    primaryClass = "hep-ph",
    reportNumber = "MITP-23-039",
    doi = "10.1007/JHEP11(2023)015",
    journal = "JHEP",
    volume = "2023",
    number = "11",
    pages = "015",
    year = "2023"
}

@article{Pustyntsev:2025nwm,
    author = "Pustyntsev, Aleksandr and Vanderhaeghen, Marc",
    title = "{Implications of recent (g-2){\ensuremath{\mu}} measurements for MeV-GeV dark sector searches}",
    eprint = "2506.17750",
    archivePrefix = "arXiv",
    primaryClass = "hep-ph",
    doi = "10.1103/b7wp-3vcj",
    journal = "Phys. Rev. D",
    volume = "112",
    number = "9",
    pages = "095001",
    year = "2025"
}

@book{Jegerlehner:2017gek,
	author = {Jegerlehner, Friedrich},
	doi = {10.1007/978-3-319-63577-4},
	title = {{The Anomalous Magnetic Moment of the Muon}},
    series = {Springer Tracts Mod.\ Phys.},
	volume = {274},
	year = {2017},
    pages={1-693}}

@article{Budapest-Marseille-Wuppertal:2017okr,
	archiveprefix = {arXiv},
	author = {Borsanyi, Sz. and others},
	collaboration = {Budapest-Marseille-Wuppertal},
	doi = {10.1103/PhysRevLett.121.022002},
	eprint = {1711.04980},
	journal = {Phys. Rev. Lett.},
	number = {2},
	pages = {022002},
	primaryclass = {hep-lat},
	title = {{Hadronic vacuum polarization contribution to the anomalous magnetic moments of leptons from first principles}},
	volume = {121},
	year = {2018}}

@article{Blokland:2001pb,
	archiveprefix = {arXiv},
	author = {Blokland, Ian Richard and Czarnecki, Andrzej and Melnikov, Kirill},
	doi = {10.1103/PhysRevLett.88.071803},
	eprint = {hep-ph/0112117},
	journal = {Phys. Rev. Lett.},
	pages = {071803},
	reportnumber = {SLAC-PUB-9084, ALBERTA-THY-16-01},
	title = {{Pion pole contribution to hadronic light by light scattering and muon anomalous magnetic moment}},
	volume = {88},
	year = {2002}}

@article{Aguillard:2025fij,
	archiveprefix = {arXiv},
	author = {Aguillard, D. P. and others},
	collaboration = {Muon g-2},
	doi = {10.1103/7clf-sm2v},
	eprint = {2506.03069},
	journal = {Phys. Rev. Lett.},
	number = {10},
	pages = {101802},
	primaryclass = {hep-ex},
	reportnumber = {FERMILAB-PUB-25-0364-PPD},
	title = {{Measurement of the Positive Muon Anomalous Magnetic Moment to 127~ppb}},
	volume = {135},
	year = {2025}}

@Article{Aliberti:2025beg,
  author        = {Aliberti, R. and others},
  journal       = {Phys. Rept.},
  title         = {The anomalous magnetic moment of the muon in the Standard Model: an update},
  year          = {2025},
  pages         = {1--158},
  volume        = {1143},
  archiveprefix = {arXiv},
  doi           = {10.1016/j.physrep.2025.08.002},
  eprint        = {2505.21476},
  primaryclass  = {hep-ph},
  reportnumber  = {CERN-TH-2025-101, FERMILAB-PUB-25-0344-T, INT-PUB-25-015, IPARCOS-UCM-25-029, KEK Preprint 2025-22, LTH 1403, MITP-25-037, UWThPh 2025-15, UWThPh 2025-15, ZU-TH 37/25, IPARCOS-UCM-25-029},
}

@article{Gourdin:1969dm,
	author = {Gourdin, M. and De Rafael, E.},
	doi = {10.1016/0550-3213(69)90333-2},
	journal = {Nucl. Phys. B},
	pages = {667--674},
	title = {{Hadronic contributions to the muon g-factor}},
	volume = {10},
	year = {1969}}

@article{Keshavarzi:2019abf,
	archiveprefix = {arXiv},
	author = {Keshavarzi, Alexander and Nomura, Daisuke and Teubner, Thomas},
	doi = {10.1103/PhysRevD.101.014029},
	eprint = {1911.00367},
	journal = {Phys. Rev.},
	number = {1},
	pages = {014029},
	primaryclass = {hep-ph},
	reportnumber = {MAN/HEP/2019/010, LTH 1216, KEK-TH-2165},
	slaccitation = {%%CITATION = ARXIV:1911.00367;%%},
	title = {{$g-2$ of charged leptons, $\alpha (M^2_Z)$ , and the hyperfine splitting of muonium}},
	volume = {D 101},
	year = {2020}}

@book{Melnikov:2006sr,
    author = "Melnikov, K. and Vainshtein, A.",
    title = "{Theory of the muon anomalous magnetic moment}",
    series = {Springer Tracts Mod.\ Phys.},
    doi = "10.1007/3-540-32807-6",
    volume = "216",
    year = "2006",
    pages={1-176}
}

@article{Melnikov:2003xd,
	archiveprefix = {arXiv},
	author = {Melnikov, Kirill and Vainshtein, Arkady},
	doi = {10.1103/PhysRevD.70.113006},
	eprint = {hep-ph/0312226},
	journal = {Phys. Rev.},
	pages = {113006},
	primaryclass = {hep-ph},
	reportnumber = {UH-511-1041-03, FTPI-MINN-03-36, UMN-TH-2224-03},
	slaccitation = {%%CITATION = HEP-PH/0312226;%%},
	title = {{Hadronic light-by-light scattering contribution to the muon anomalous magnetic moment revisited}},
	volume = {D 70},
	year = {2004}}

@article{Antognini:2022xoo,
	archiveprefix = {arXiv},
	author = {Antognini, Aldo and Hagelstein, Franziska and Pascalutsa, Vladimir},
	doi = {10.1146/annurev-nucl-101920-024709},
	eprint = {2205.10076},
	journal = {Ann. Rev. Nucl. Part. Sci.},
	pages = {389},
	primaryclass = {nucl-th},
	reportnumber = {MITP/22-039, PSI-PR-22-29},
	title = {{The proton structure in and out of muonic hydrogen}},
	volume = {72},
	year = {2022}}

@article{Parker:2018vye,
	archiveprefix = {arXiv},
	author = {Parker, Richard H. and Yu, Chenghui and Zhong, Weicheng and Estey, Brian and M{\"u}ller, Holger},
	doi = {10.1126/science.aap7706},
	eprint = {1812.04130},
	journal = {Science},
	pages = {191},
	primaryclass = {physics.atom-ph},
	slaccitation = {%%CITATION = ARXIV:1812.04130;%%},
	title = {{Measurement of the fine-structure constant as a test of the Standard Model}},
	volume = {360},
	year = {2018}}

@book{Peskin:1995ev,
	author = {Peskin, Michael E. and Schroeder, D. V.},
	publisher = {Addison-Wesley},
	title = {{An Introduction to quantum field theory}},
	year = {1995}}

@article{Lautrup:1968tdb,
	author = {Lautrup, B. E. and De Rafael, E.},
	doi = {10.1103/PhysRev.174.1835},
	journal = {Phys. Rev.},
	pages = {1835--1842},
	title = {{Calculation of the sixth-order contribution from the fourth-order vacuum polarization to the difference of the anomalous magnetic moments of muon and electron}},
	volume = {174},
	year = {1968}}

@article{bijnens:2019ghy,
	archiveprefix = {arXiv},
	author = {Bijnens, Johan and Hermansson-Truedsson, Nils and Rodr{\'\i}guez-S{\'a}nchez, Antonio},
	doi = {10.1016/j.physletb.2019.134994},
	eprint = {1908.03331},
	journal = {Phys. Lett. B},
	pages = {134994},
	primaryclass = {hep-ph},
	reportnumber = {LU TP 19-38},
	slaccitation = {%%CITATION = ARXIV:1908.03331;%%},
	title = {{Short-distance constraints for the HLbL contribution to the muon anomalous magnetic moment}},
	volume = {798},
	year = {2019}}

@article{Knecht:2001qg,
	archiveprefix = {arXiv},
	author = {Knecht, M. and Nyffeler, Andreas and Perrottet, M. and de Rafael, E.},
	doi = {10.1103/PhysRevLett.88.071802},
	eprint = {hep-ph/0111059},
	journal = {Phys. Rev. Lett.},
	pages = {071802},
	primaryclass = {hep-ph},
	reportnumber = {CPR-2001-P-4260},
	slaccitation = {%%CITATION = HEP-PH/0111059;%%},
	title = {{Hadronic light by light scattering contribution to the muon g-2: An Effective field theory approach}},
	volume = {88},
	year = {2002}}

@article{Biloshytskyi:2025fjm,
    author = "Biloshytskyi, Volodymyr and Erb, Dominik and Meyer, Harvey B. and Parrino, Julian and Pascalutsa, Vladimir",
    title = "{Field-theoretic versus data-driven evaluations of electromagnetic corrections to hadronic vacuum polarization in $(g-2)_\mu $}",
    eprint = "2509.08115",
    archivePrefix = "arXiv",
    primaryClass = "hep-ph",
    reportNumber = "CERN-TH-2025-184, MITP-25-059",
    doi = "10.1140/epjc/s10052-026-15521-6",
    journal = "Eur. Phys. J. C",
    volume = "86",
    number = "5",
    pages = "497",
    year = "2026"
}

@article{Morel:2020dww,
    author = {Morel, L{\'e}o and Yao, Zhibin and Clad{\'e}, Pierre and Guellati-Kh{\'e}lifa, Sa{\"\i}da},
    title = "{Determination of the fine-structure constant with an accuracy of 81 parts per trillion}",
    doi = "10.1038/s41586-020-2964-7",
    journal = "Nature",
    volume = "588",
    number = "7836",
    pages = "61--65",
    year = "2020"
}

@article{Bouchiat:1961lbg,
    author = "Bouchiat, Claude and Michel, Louis",
    title = "{La r{\'e}sonance dans la diffusion m{\'e}son {\ensuremath{\pi}}{\textemdash} m{\'e}son {\ensuremath{\pi}} et le moment magn{\'e}tique anormal du m{\'e}son {\ensuremath{\mu}}}",
    doi = "10.1051/jphysrad:01961002202012101",
    journal = "J. Phys. Radium",
    volume = "22",
    number = "2",
    pages = "121--121",
    year = "1961"
}

@article{PhysRev.168.1620,
  title = {Suggested Boson-Lepton Pair Couplings and the Anomalous Magnetic Moment of the Muon},
  author = {Brodsky, Stanley J. and de Rafael, Eduardo},
  journal = {Phys. Rev.},
  volume = {168},
  issue = {5},
  pages = {1620--1622},
  numpages = {0},
  year = {1968},
  month = {Apr},
  publisher = {American Physical Society},
  doi = {10.1103/PhysRev.168.1620},
  url = {https://link.aps.org/doi/10.1103/PhysRev.168.1620}
}

@article{Fan:2022eto,
    author = "Fan, X. and Myers, T. G. and Sukra, B. A. D. and Gabrielse, G.",
    title = "{Measurement of the Electron Magnetic Moment}",
    eprint = "2209.13084",
    archivePrefix = "arXiv",
    primaryClass = "physics.atom-ph",
    doi = "10.1103/PhysRevLett.130.071801",
    journal = "Phys. Rev. Lett.",
    volume = "130",
    number = "7",
    pages = "071801",
    year = "2023"
}

@article{Ilten:2018crw,
    author = "Ilten, Philip and Soreq, Yotam and Williams, Mike and Xue, Wei",
    title = "{Serendipity in dark photon searches}",
    eprint = "1801.04847",
    archivePrefix = "arXiv",
    primaryClass = "hep-ph",
    reportNumber = "MIT-CTP/4976, CERN-TH-2017-282, MIT-CTP-4976",
    doi = "10.1007/JHEP06(2018)004",
    journal = "JHEP",
    volume = "2018",
    number = "06",
    pages = "004",
    year = "2018"
}

@article{FASER:2023tle,
    author = "Abreu, Henso and others",
    collaboration = "FASER",
    title = "{Search for dark photons with the FASER detector at the LHC}",
    eprint = "2308.05587",
    archivePrefix = "arXiv",
    primaryClass = "hep-ex",
    reportNumber = "CERN-EP-2023-161",
    doi = "10.1016/j.physletb.2023.138378",
    journal = "Phys. Lett. B",
    volume = "848",
    pages = "138378",
    year = "2024"
}

\end{document}